\documentclass[12pt,preprint]{aastex}

\begin{document}

\title{A New Solution for the Dispersive Element in Astronomical Spectrographs}
 
\author{Harland W. Epps\altaffilmark{1}, Judith G. Cohen\altaffilmark{2}
\&  J. Christopher Clemens\altaffilmark{3} }

\altaffiltext{1}{UCO/Lick Observatory, University of California,
Santa Cruz, Ca., 95064, epps@ucolick.org}

\altaffiltext{2}{Palomar Observatory, Mail Stop 249-17,
California Institute of Technology, Pasadena, Ca., 91125,
jlc@astro.caltech.edu}

\altaffiltext{3}{Department of Physics and Astronomy, University of
North Carolina, Chapel Hill, NC, 27599-3255,
clemens@physics.unc.edu}


\begin{abstract}
We present a new solution for the dispersive
element in astronomical spectrographs,
which  in many cases can provide an 
upgrade path to enhance the spectral
resolution of existing moderate-resolution reflection-grating 
spectrographs.  We demonstrate that in the case of LRIS-R
at the Keck~1 Telescope a spectral resolution of 18,000
can be achieved with reasonable throughput under good
seeing conditions.

\end{abstract}

\section{Introduction}

We have been carrying out a preliminary design study to explore
achieving multiplexing 
of high-resolution optical spectroscopy
for point sources  at the Keck
Observatory.   This effort arose from our desire to provide
capabilities for the Keck user community comparable
to those offered by the successful fiber-optic-fed
instruments
FLAMES+GIRAFFE and FLAMES+UVES (Pasquini et al. 2002, Dekker et al. 2000).
These multiplex up to 130 targets to the medium-high-resolution optical
GIRAFFE spectrograph with up to 8 targets to
the VLT's high-resolution UVES spectrograph.  These instruments
have made major advances possible in a number of
areas including near-field cosmology and Galactic stellar spectroscopy.
Since the cost of building a new instrument for the Keck Observatory
for this purpose is in excess of 15 million dollars, we 
explored the possibility of modifying an existing 
moderate-dispersion multi-object spectrograph to achieve this goal in a way that also 
preserves the present mode of operation of the spectrograph.

In the course of this effort, we found what
we believe to be a new solution for the dispersive
element in astronomical spectrographs.  It 
substantially enhances the spectral resolution
of suitable existing moderate-resolution spectrographs. 
Our solution can thus
provide a path for upgrading and augmenting the capability of many
existing instruments.  It will be economical
and, in most cases, fairly easy to implement.
In the case of new spectrographs,
we believe adopting this solution for the disperser, instead of a conventional
reflection grating,
can result in the design of 
smaller, cheaper instruments that can achieve a given specified
high spectral resolution. 

\section{Existing Solutions \label{section_exist}}

Some efforts to increase the spectral resolution of an
existing multi-object spectrograph have involved inserting an echelle
grating and using the existing reflection grating (or
a prism) as a cross disperser.  
This concept was included in the
design of the Norris Spectrograph for the 5-m Hale
Telescope at Palomar Mountain \citep{hamilton_norris}
but the echelle mode was never implemented, due to lack of funds.
To maintain object multiplexing in such a spectrograph,
the number of echelle orders to be used
must be limited by interference filters to be a small number.
 
The multi-object high-resolution
echellete addition to IMACS on the Magellan 6.5-m Telescope
is a  variant of this scheme; a prism cross-dispersed echellete grating
is now available as an optional module on the moderate-resolution IMACS
instrument \citep{sutin_moe}.  The echellette requires
10 orders for full spectral coverage over
the optical regime; a filter can be used to restrict
the wavelength range retained and hence to increase the multiplexing
capability.

Unfortunately as the telescope diameter increases
beyond that of the 6.5-m Magellan Telescopes, this upgrade
path  becomes more difficult and we have established that it
is not  practical
for the two Keck multi-object moderate-resolution
optical spectrographs,
LRIS \citep{oke_lris} and DEIMOS \citep{faber_deimos}.

Other techniques that can be used to
try to achieve higher spectral resolution include
increasing the line density or
using a reflection grating
nominally designed for use in the first order  near IR,
to cover  optical wavelengths in the second order instead.
While this doubles the spectral dispersion, it  also
leads to very large angles of incidence and large anamorphic
factors.  The anamorphism reduces the projected slit
width, which can lead to inadequate pixel sampling.
The large angle of incidence and hence 
large projection factor also demands a grating with one dimension much
larger than the beam size, which is typically
$\sim$150~mm in many optical spectrographs.  
If the grating does not have the required width in 
the direction of the dispersion, a severe vignetting
light loss will occur.  A suite of 154 x 206 mm gratings
is available from the Richardson Grating Laboratory
of the Newport Corporation, but larger ones are not readily
available.  Large anamorphic factors also place added
demands on the camera's entrance aperture and optical
design, which are not normally satisfied in
moderate-resolution spectrographs.

High quality reflective gratings with rulings exceeding 1200 lines/mm,
suitable for first order use at optical
wavelengths and in sizes large enough for a 150-mm
diameter beam, are not commercially available to
our knowledge.  Ruling a new master for such a high-resolution
grating would be very expensive, and would require
a minimum timescale of a year.  There is only
one commercial source for such a 
large astronomical grating in the United States
today, the Richardson Grating Laboratory, now part of
Newport Corporation.

\section{Our New Solution}

Our solution involves replacing the normal low-order reflection
grating with a {\it{pair}} of volume phase holographic (VPH)
gratings used in transmission.
The use of a pair of VPH gratings enables bending the beam
so that the axial light path matches
that of a normal reflection-grating spectrograph
with an included angle $Q$ between the collimator and camera
optical axes.
VPH gratings, unlike reflection gratings,
are currently available  from several commercial vendors
in the required (large) sizes,
with much higher line densities, for a price less
than \$20,000 each.  The timescale for
acquiring such VPH gratings is considerably shorter than that
for ruling a new master for a large reflection grating.

\begin{figure}
\epsscale{1.0}
\plotone{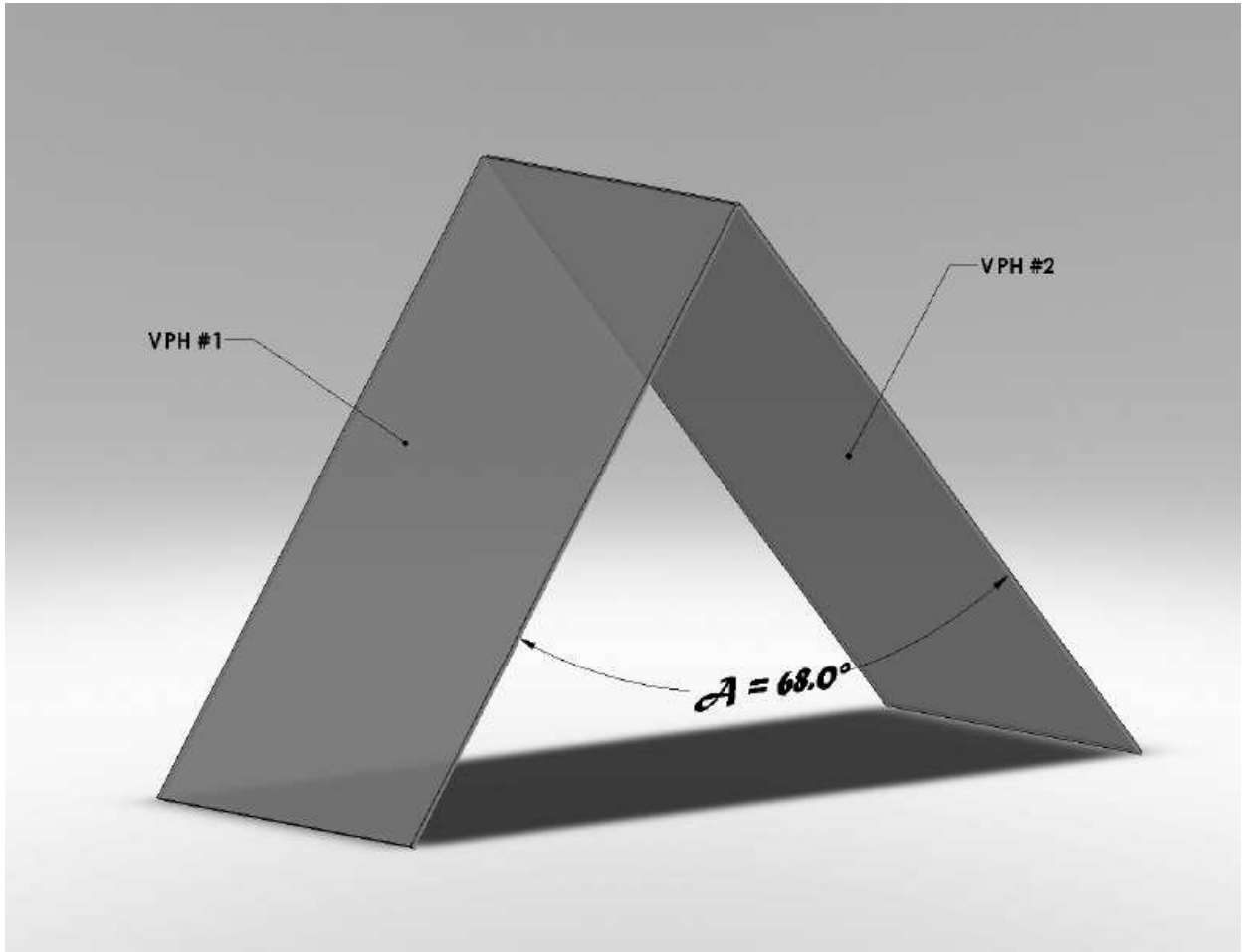}
\caption[]{A sketch of a pair of VPH gratings assembled into a ``tent'' to function
as a disperser is shown.
\label{figure_tent}}
\end{figure}

The two identical VPH gratings in the pair 
form a ``tent'', with both ends open, whose
entrance and exit are identical isoceles triangles and whose sides
are each a VPH grating as shown in Fig.~\ref{figure_tent}.  
The apex angle of the isoceles
triangle ($A$) is fixed at a value of ($90.0 - (Q/2)$) so that light from the 
collimator at the
desired central wavelength
passes through the VPH pair and ends up heading toward
the camera when the diffraction of each VPH is taken into account.
If the VPH pair is to be 
used to upgrade an existing spectrograph, to allow a higher-resolution mode, 
the included angle $Q$ between the collimator and camera optical
axes is fixed and is 
determined by the design of the existing instrument.

\begin{figure}
\epsscale{1.0}
\plotone{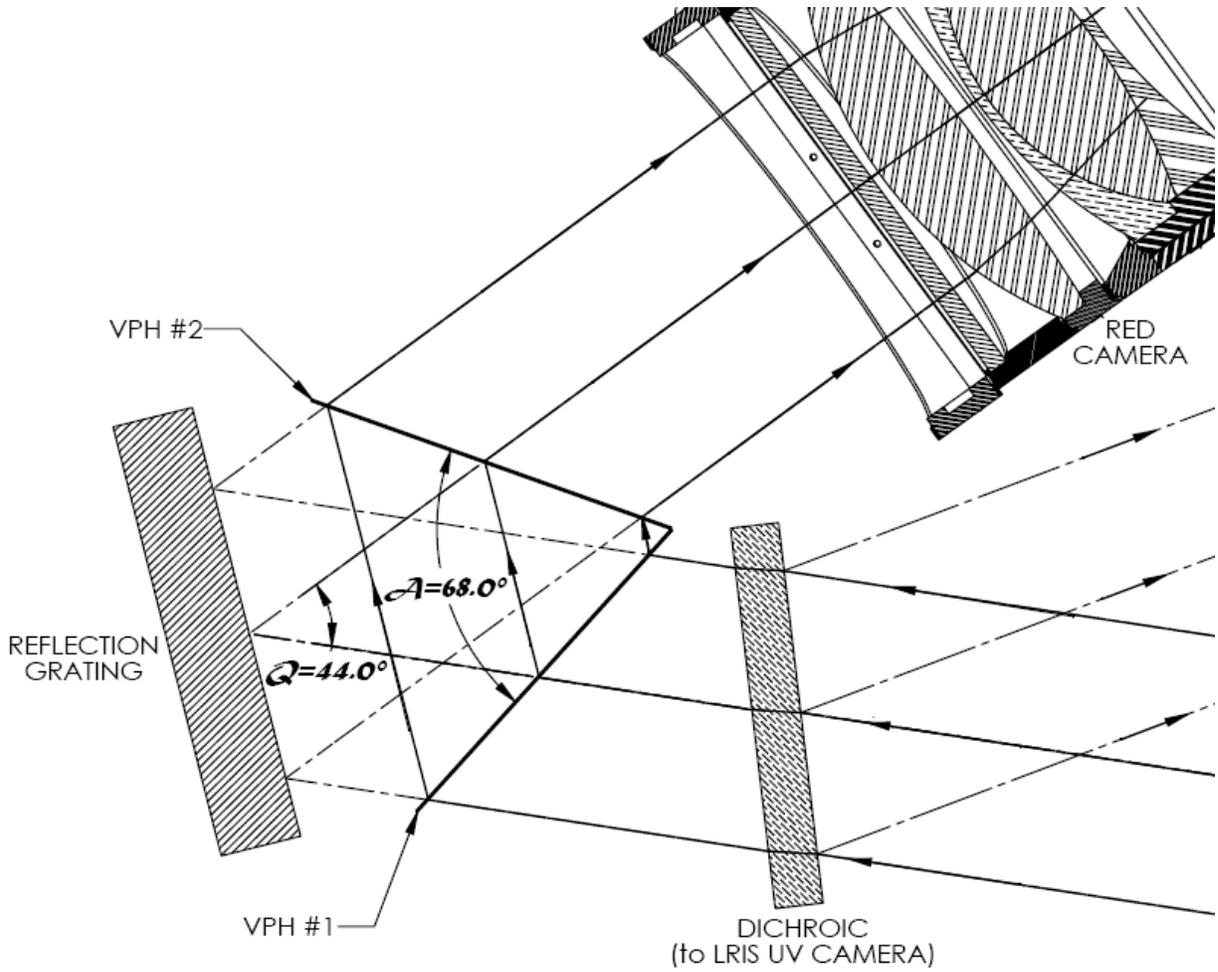}
\caption[]{The region around the reflection grating used as a disperser
in the moderate-resolution LRIS spectrograph at the Keck~1 Telescope
is shown.  The placement of the proposed VPH pair upgrade
to achieve higher spectral resolution while maintaining
the multiplexing capability of LRIS-R is also indicated.
\label{figure_lris}}
\end{figure}

Once the apex angle $A$ is known, there is only one free design
parameter for the dispersive element, 
the line density of the VPH grating pair, $N_l$ lines/mm.  
Given the fixed included angle $Q$, the dispersion (in radians/$\mu$)
is  proportional to $N_l$ of the VPH gratings
while the central wavelength is  proportional to $N_l^{-1}$. 
For a given slit width, the spectral resolution is independent
of $N_l$.
The spectral coverage
is then limited, when this is used as an upgrade to an
existing instrument, by the width of the detector, which sets
the extent of the first order that fits across the detector.
Fig.~\ref{figure_lris} shows a layout for a VPH pair in the
Red arm of the  existing LRIS
spectrograph at the Keck~1 Telescope.  Some performance quantities
are tabulated in Table~\ref{table_lris} for the double-VPH disperser 
mode in an upgraded LRIS-R.

VPH gratings to accommodate a $\sim$150-mm beam can  be manufactured
with $N_l \sim 3000$~lines/mm, allowing the possibility
of reaching much higher spectral resolution than in a normal
first-order reflection grating spectrograph, for the same projected
slit width.  The gain in spectral resolution over a 1200 lines/mm reflection
grating used in first order in a normal astronomical spectrograph is then
$\approx 2 {N_l / [1200 ~ Cos(A/2)]}$ when a VPH pair is used
as described above.

\subsection{Advantages}

The advantages of the proposed
double-VPH disperser are many.  There are no moving or tiltable
parts. Mechanical flexure, the bane of many existing spectrographs,
is easier to control as a VPH pair is considerably lighter
than a reflection grating for the same beam size and the
former does not have to be rotated.
VPH gratings, at least up to sizes
of $\sim$205~mm on a side, are readily available at ``reasonable''
prices.  

If the existing spectrograph to be modified has multi-object
capability, which is  more common in moderate-resolution
instruments than in high-resolution ones, this capability
is unaffected by the introduction of the double-VPH grating
as the dispersive element; the instrument will perform in the same
way when the dispersive element is changed to the VPH pair,
making this an ideal upgrade path.  

Furthermore, adding
the option of a VPH pair to a moderate-resolution multi-object
spectrograph enables the use of both modes, provided that a switch
between the two types of dispersers is implemented.  Thus
both can take advantage of the multiplexing already incorporated
into the  moderate-resolution instrument.

For a particular pair of VPH gratings,
the central wavelength and
spectral range in each order are fixed by $N_l$ once
the included angle $Q$ between the collimator and camera
optical axes is
specified, either in a new design or an as-built
design. 
Given the relatively short wavelength regime
a given VPH pair must cover, it is not difficult
to make their throughput comparable for the pair to
that of a conventional reflection grating intended for use
over a much broader wavelength regime.  Quantitative details
are given in \S\ref{section_efficiency}.

In the design of
astronomical spectrographs, aberrations must be balanced,
and the
final image often suffers from some degree of axial chromatic
aberration.  Since  the
spectral range within which a VPH pair will be used
is relatively small (limited by the detector width),
the axial chromatic
aberration can largely be eliminated for a given
setup by optimizing the spectrograph focus for each
VPH pair.

\subsection{Disadvantages}

There are of course disadvantages as well.  The major
 one is the limited spectral range that can be covered by
 a given VPH pair due to the fixed collimator to camera 
 included angle $Q$.  Given this,
several such VPH  pairs will need to be purchased, each of which
covers a spectral window of particular interest.  Such
a scheme is best suited for stellar work where
the redshift is fixed at $z \sim 0.0$.
The variety of redshifts of extragalactic sources
makes it impossible to optimize
and limit the number of wavelength ranges that must
be covered.  Full coverage of redshift space using VPH gratings
would be costly, but not nearly as expensive as building a new
high-resolution multiplexing spectrograph from scratch.

Ideally the required pairs of VPH
gratings for most/all observations
can all be accommodated within a slide
or storage box plus changer mechanism designed
with the safety of the instrument optics as well
as ease of use in mind.  This would make it possible for
all the VPH grating pairs that are to be used in a given observing
run to be 
installed within the instrument, and the pair in use
could be switched during the night 
via remote control.  If there are more VPH pairs in
regular use than can be accommodated by such a mechanism,
manual switching would be necessary. This is undesirable
as such manual changes
would become a maintenance issue and affect the long-term
operational cost.  There is also a concern for the
safety of the other optical components in the spectrograph
during any procedure that requires manual access to its interior.

Implementation of the VPH pair within an existing instrument
requires that the area around the reflection grating
be fairly open and accessible so that a holder
for the VPH grating pairs can be installed,
ideally with a changer mechanism.  This mechanism might
be as simple as a manually (or better yet a remotely-driven)
slide of VPH
pairs.   Some existing spectrographs
contain other components, which occupy the
desired volume for other key purposes, making
it difficult or impossible to implement
double-VPH gratings as an upgrade path.
In addition, there is
a requirement that the volume to be occupied by
the VPH pair, which extends  toward both
the collimator and the camera from the nominal
position of a reflection-grating disperser (see Fig.~\ref{figure_lris}), 
does not include any
part of the optical path that would 
obscure the beam in part or completely.

\section{Predicted LRIS-R Performance Comparison \label{section_lris_red}}

The LRIS-R camera has recently been upgraded with two
(2048 by 4096 by 15-$\mu$) fully-depleted high-resistivity thick red-sensitive
CCD's from Lawrence Berkeley Laboratory \citep{stover_ccd}.
The measured LRIS-R final image scale on the Keck~1 Telescope
  is 0.135 arcsec/pixel \citep{rockosi_scale},  which is consistent 
with the camera's original 305.0~mm
effective focal length \citep{oke_lris} even though a new field flattener
lens was installed during the upgrade.

     A 1200 lines/mm reflection grating working in first order at a central
wavelength of 5886.2~\AA\ is selected for 
detailed comparison with the 1900~lines/mm
double-VPH example in Table~\ref{table_lris}.  The reflection grating set up produces a rather
extreme anamorphic factor ($r = 1.3993$) and a finer ruling is not practical for
reasons described in Section~\ref{section_exist} above.  A larger anamorphic factor would cause
excessive vignetting and would lead to inadequate image quality as it would
overdrive the camera's optical design.  Thus near-maximum reflection-grating
resolution is expected in this example.

     The comparison must be made with the same effective slit width(s)
in {\it{pixels}} 
at the CCD, for both modes.  Anamorphic demagnification will enable the slit
to be wider in {\it{arcsec}} by the
anamorphic factor ($r$) relative to
the double-VPH mode, which has no anamorphism.  The quantitative comparison is
given in Table~\ref{table_comp}.

     Table~\ref{table_comp} is not extended further because the spectral resolution would
tend to become pixel-sampling limited;  residual aberrations in the camera
would limit the resolution as well.  The effective silt-width limit should
be determined experimentally but practical experience suggests that about
3.0~pixels will be a close approximation.

     At a given pixel-sampling, the wider slit in reflection-grating mode
will tend to favor its throughput.  However that advantage will tend to
disappear with better ``seeing'', and vignetting at the camera's entrance
aperture due to the large anamorphism will cause light loss toward the
ends of the spectrum as well.  Thus we assert that it will be quite feasible
to produce photon-efficient multiplexed spectra at resolutions of 
$R = 18,000$ or more with our
proposed double-VPH disperser(s), used in LRIS-R with good ``seeing''.  The
fact that double-VPH resolution is independent of central wavelength is
an important added advantage not shared by the reflection-grating mode.

\subsection{Efficiency of the Double-VPH Disperser \label{section_efficiency}}

\begin{figure}
\epsscale{1.0}
\plotone{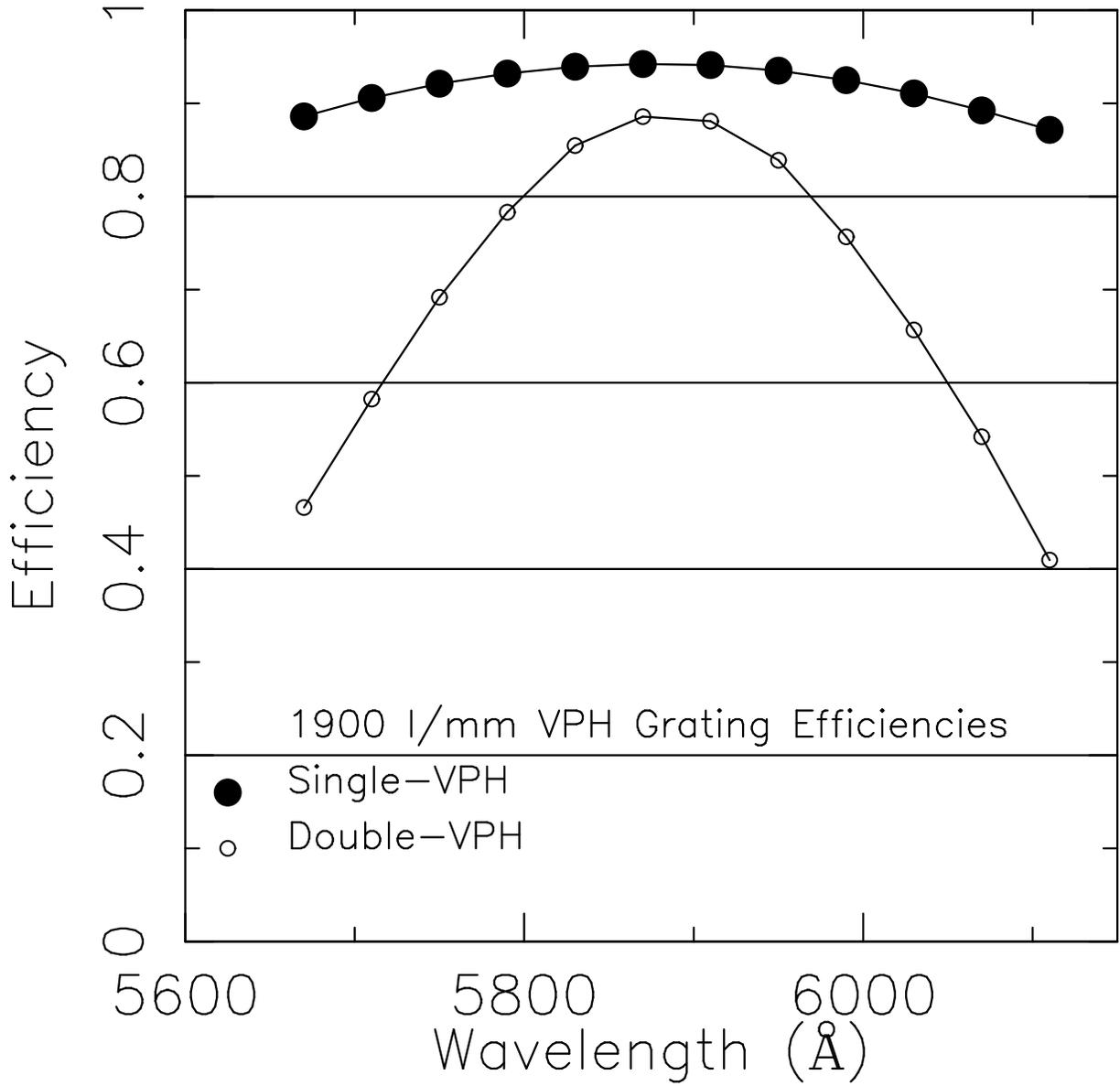}
\caption[]{The efficiency of a single 1900 l/mm VPH grating
is shown, as is that of a pair of them
designed for use in LRIS-R.  The wavelength coverage here and in the
following Figures is limited by the CCD detector width.
\label{figure_dp1900}}
\end{figure}

We have used rigorous coupled wave analysis (RCWA) to investigate
the efficiency of VPH gratings used in a double-VPH mode.  
The Fresnel losses at the four air/glass boundaries have been neglected.
In total they would be $\sim$1\% depending on the details of
the coatings over each of the narrow passbands.
Fig.~\ref{figure_dp1900} shows calculations
for a pair of 1900 l/mm gratings optimized for use at an
incident angle of 34$^{\circ}$, which is appropriate for LRIS-R.
The film thickness used in this model is 4.5~$\mu$, and the index
modulation is 0.07, both of which are easily achievable by any of the
VPH grating vendors.  The upper curve shows the efficiency ($\epsilon$) 
in single-pass, over the spectral range that is covered by the recently
upgraded 
LRIS-R detector. Light at the Bragg condition (5886~\AA\ for this
line density) is diffracted most efficiently and hits the second
grating also at the Bragg condition, so the total efficiency
is simply $\epsilon^2$.  However, incident
light at wavelengths different from the Bragg condition is
not only diffracted less efficiently by the first grating, but
also strikes the second grating even further 
from the Bragg condition because of the angular deviation
introduced by the dispersive effect of the first grating.  The penalty
for this angular deviation is higher than for the mismatch to the
Bragg wavelength, so the efficiency drops more steeply than 
$\epsilon^2$ at the off-Bragg wavelengths, as illustrated by
the lower curve.  The efficiency averaged over the
440~\AA\ wide spectral range
is still quite high, $\sim 70\%$.

\begin{figure}
\epsscale{1.0}
\plotone{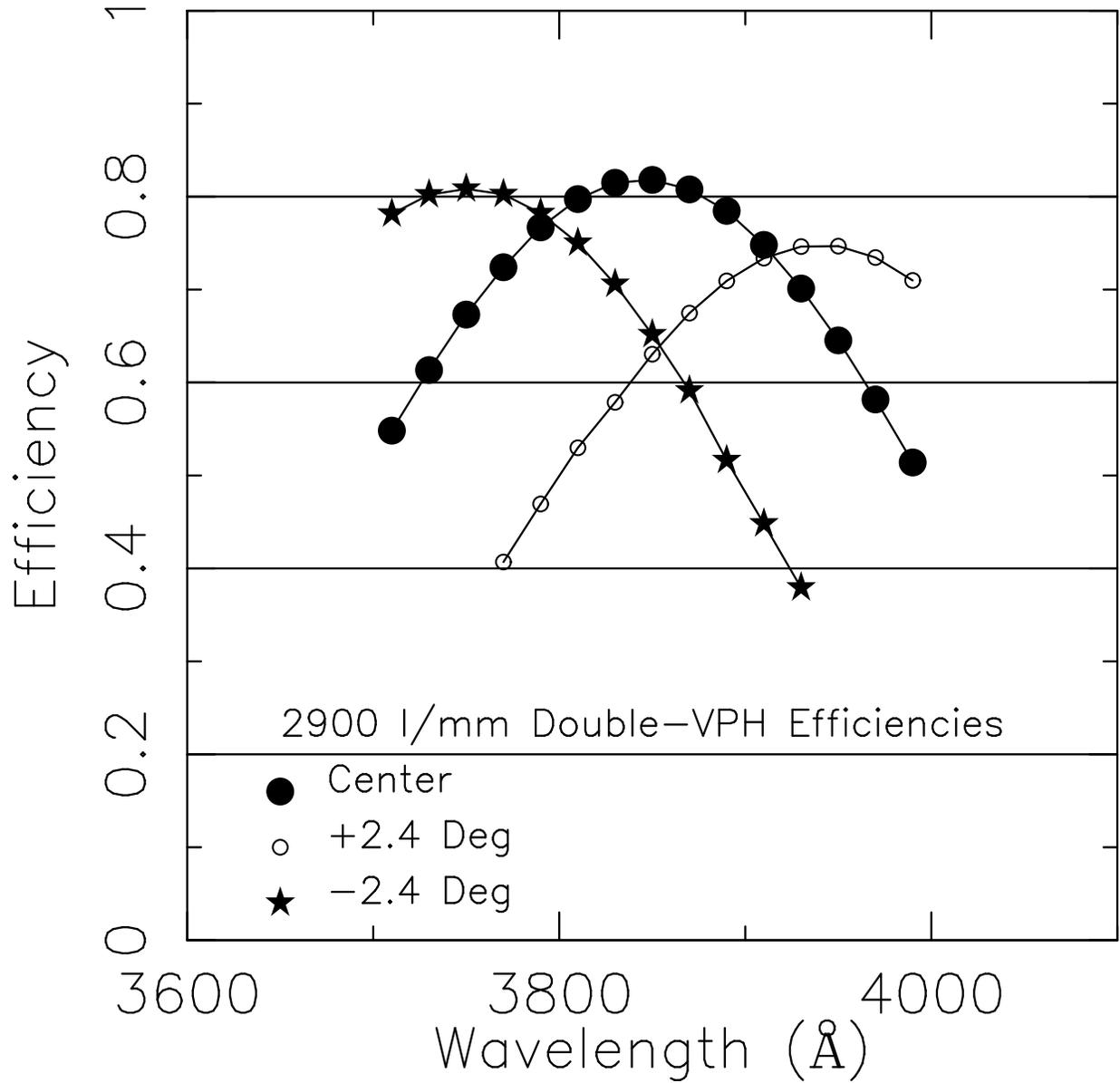}
\caption[]{The efficiency of a pair of VPH gratings with
2900~l/mm, 2.5~$\mu$ thickness, and 0.07 index modulation
is shown.  Curves are also shown for efficiencies and 
wavelength coverage for objects that
illuminate the first grating at $\pm 2.4^{\circ}$ from the 34$^{\circ}$
central angle, as happens in multi-slit mode extreme cases.
\label{figure_dp2900}}
\end{figure}

\begin{figure}
\epsscale{1.0}
\plotone{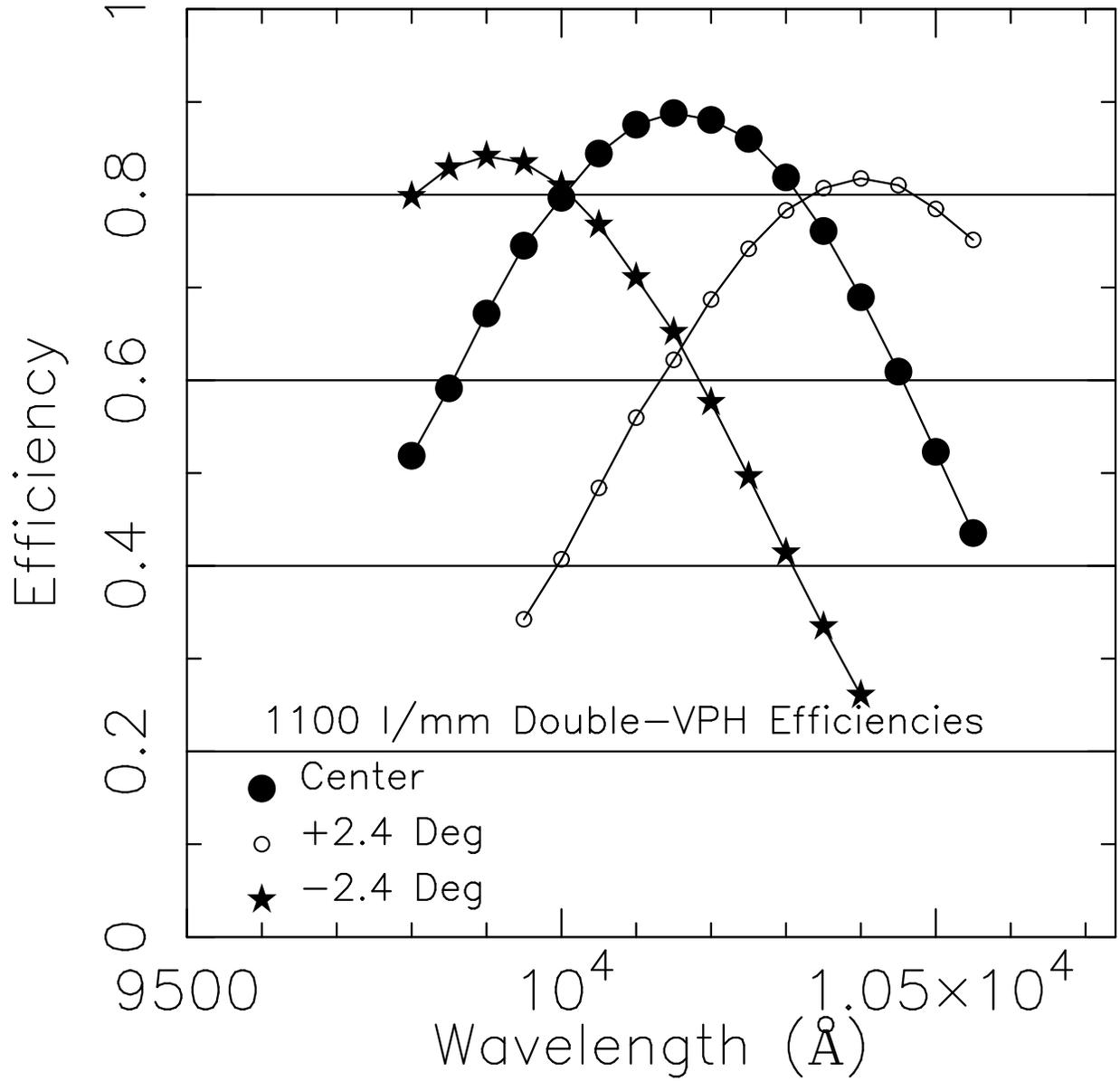}
\caption[]{The same as Fig.~\ref{figure_dp2900}, but for a pair of
VPH gratings with 1100~l/mm, 7.5~$\mu$ thickness, and 0.07 index
modulation.
\label{figure_dp1100}}
\end{figure}

In addition to understanding the efficiency of the 1900 l/mm grating, 
we have used RCWA to investigate whether the entire grating
complement listed in Table~\ref{table_lris} can be made
at similar high efficiencies.  We have modeled a family of gratings,
all of which have 0.07 index modulation, but with thicknesses
that vary more-or-less in inverse proportion to the
line density.  Figures~\ref{figure_dp2900} and 
\ref{figure_dp1100} show the double-VPH efficiency curves
for the highest and lowest line density grating models, also
over the wavelength range that can be detected in each case 
with LRIS-R.  The most difficult grating to fabricate is that with
2900 l/mm because it is only 2.5~$\mu$ thick.  \cite{blanche04} have
published confirmation that CSI/ATHOL can make efficient
gratings  of 2.9~$\mu$ thickness, but they have not confirmed
this for high line density.  However during the preparation of this
paper, we were able to produce a sample in the Goodman Laboratory
that has 3100~l/mm, a thickness of 2.5~$\mu$, and a modulation
of 0.07, nearly identical to the grating modeled in 
Fig.~\ref{figure_dp2900}.

It is also important to model $\epsilon$ for incident light
that strikes the first grating off-axis.  This can occur because
LRIS is a multi-object spectrograph.  Slitlets 
at extreme angles away from the
centerline of the slit mask in the dispersion  direction,
illuminate the grating at angles $\pm 2.4^{\circ}$
from the central angle.  Figures~\ref{figure_dp2900} and 
\ref{figure_dp1100} include curves showing this effect for slitlets
displaced at these extreme locations.  The peak 
efficiency is shifted from that of the on-axis curve and 
the mean efficiency is somewhat lower, although it averages
$\sim 65$\% as shown. The spectral coverage
is shifted as well, depending on the distance of the slitlet
off of the centerline of the slit mask, an 
effect which always occurs in multi-slit
spectrographs.

We conclude from these representative calculations that
double-VPH dispersers can be expected to provide average
efficiencies of $\sim$ 65 to 70\% over the entire complement
of grating pairs and the full range of multi-slit locations
available to LRIS-R.

\acknowledgements

The authors acknowledge partial support from
a grant of seed money from the Keck Observatory to 
J.Cohen, H.Epps and M. Rich.  We
thank Michael Rich for his long term advocacy for multiplexed
high resolution spectroscopy at the Keck Observatory.

{}

\clearpage

\begin{deluxetable}{l rrrr}
\tablenum{1}
\tablewidth{0pt}
\tablecaption{Properties of VPH Pairs In LRIS-R at the 10-m Keck~1 Telescope
\label{table_lris}}
\tablehead{
\colhead{Lines/mm} & \colhead{Central $\lambda$} & \colhead{Dispersion} &
\colhead{Spectral coverage (\AA)} \\
\colhead{of VPH pair} & \colhead{(\AA)} & 
\colhead{(radians/$\mu$)} & \colhead{(LRIS-R at Keck)\tablenotemark{a}} 
}
\startdata
1100  &  10167 & 2.6537 & 759 \\
1200  &   9320 & 2.8949 & 696 \\
1300  &   8603 & 3.1362 & 642 \\
1500  &   7456 & 3.6187 & 557 \\
1700  &   6579 & 4.1011 & 491 \\
1900  &   5886 & 4.5836 & 439 \\
2100  &   5326 & 5.0661 & 398 \\
2300  &   4863 & 5.5486 & 363 \\
2500  &   4474 & 6.0311 & 334 \\
2700  &   4142 & 6.5136 & 309 \\
2900  &   3857 & 6.9961 & 288 \\
\enddata
\tablenotetext{a}{Section~\ref{section_lris_red} gives the relevant 
characteristics of LRIS-R and its detector.}
\end{deluxetable}

\begin{deluxetable}{l | rl | rr}
\tablenum{2}
\tablewidth{0pt}
\tablecaption{Resolution Comparison for Optimal Reflection-Grating Mode
          vs. Double-VPH Dispersion Mode in LRIS-R.
\label{table_comp}}
\tablehead{
\colhead{Slit Width} & \colhead{Refl. Grat.} & \colhead{Resolution} &
\colhead{Double-VPH} & \colhead{Resolution} \\
\colhead{(pixels)} & \colhead{(arcsec)} & 
\colhead{} & \colhead{(arcsec)} 
}
\startdata
 7.0 & 1.32 & 2,052 & 0.95 & 7,837 \\
 6.0 & 1.13 & 2,394 & 0.81 & 9,143 \\
 5.0 & 0.94 & 2,872 & 0.68 & 10,972 \\
 4.0 & 0.76 & 3,591 & 0.54 & 13,715 \\
 3.0 & 0.57 & 4,787 & 0.41 & 18,287 \\
\enddata
\end{deluxetable}

\clearpage

\end{document}